\documentclass[twocolumn]{aastex631}

\def\specialname[#1]{\textbf{\textsc{#1}}}

\usepackage{multirow}

\shorttitle{Environmental dependence of the mass-metallicity relation}
\shortauthors{Wang et al.}
\graphicspath{{./}{figures/}}

\begin{document}

\title{Environmental dependence of the mass-metallicity relation in cosmological hydrodynamical simulations}

\author[0000-0002-3775-0484]{Kai Wang}
\affiliation{\rm Kavli Institute for Astronomy and Astrophysics, Peking University, Beijing 100871, China; \href{mailto:wkcosmology@gmail.com}{wkcosmology@gmail.com}}

\author[0000-0002-9373-3865]{Xin Wang}
\affiliation{\rm School of Astronomy and Space Science, University of Chinese Academy of Sciences (UCAS), Beijing 100049, China; \href{mailto:xwang@ucas.ac.cn}{xwang@ucas.ac.cn}}
\affiliation{\rm National Astronomical Observatories, Chinese Academy of Sciences, Beijing 100101, China}
\affiliation{\rm Institute for Frontiers in Astronomy and Astrophysics, Beijing Normal University, Beijing 102206, China}

\author[0000-0002-4597-5798]{Yangyao Chen}
\affiliation{\rm School of Astronomy and Space Science, University of Science and Technology of China, Hefei 230026, China}
\affiliation{\rm Key Laboratory for Research in Galaxies and Cosmology, Department of Astronomy, University of Science and Technology of China, Hefei, 30026, China}



\begin{abstract}
    We investigate the environmental dependence of the gas-phase metallicity
    for galaxies at $z=0$ to $z\gtrsim 2$ and the underlying physical
    mechanisms driving this dependence using state-of-the-art cosmological
    hydrodynamical simulations. We find that, at fixed stellar mass, central
    galaxies in massive halos have lower gas-phase metallicity than those in
    low-mass halos. On the contrary, satellite galaxies residing in more
    massive halos are more metal-rich. The combined effect is that massive
    galaxies are more metal-poor in massive halos, and low-mass galaxies are
    more metal-rich in massive halos. By inspecting the environmental
    dependence of other galaxy properties, we identify that the accretion of
    low-metallicity gas is responsible for the environmental dependence of
    central galaxies at high $z$, whereas the AGN feedback processes play a
    crucial role at low $z$. For satellite galaxies, we find that both the
    suppression of gas accretion and the stripping of existing gas are
    responsible for their environmental dependence, with negligible effect from
    the AGN feedback. Finally, we show that the difference of gas-phase
    metallicity as a function of stellar mass between protocluster and field
    galaxies agrees with recent observational results, for example from the
    MAMMOTH-Grism survey.
\end{abstract}



\section{Introduction}%
\label{sec:introduction}

Gas-phase metallicity is an important probe of a variety of astrophysical
processes, like gas inflow, outflow, and star formation activity, in galaxy
formation and evolution. In the past decades, several scaling relations for the
gas-phase metallicity have been established based on the galaxy survey data
from our local volume to high-$z$ Universe, like the mass-metallicity relation
(MZR) \citep[e.g.][]{tremontiOriginMassMetallicityRelation2004,
erbMassMetallicityRelation2006, maiolinoMOONRISEMainMOONS2020} and the
fundamental metallicity relation (FMR)
\citep[e.g.][]{ellisonCluesOriginMassMetallicity2008,
mannucciFundamentalRelationMass2010, dayalPhysicsFundamentalMetallicity2013}.
Meanwhile, theoretical models and numerical simulations have provided us
insights into the underlying physical processes that drive these scaling
relations \citep[e.g.][]{boucheImpactColdGas2010,
    dayalPhysicsFundamentalMetallicity2013, lillyGASREGULATIONGALAXIES2013,
    dekelAnalyticSolutionMinimal2014, pengHaloesGalaxiesDynamics2014,
kacprzakColdmodeAccretionDriving2016}.

Galaxy evolution is not only affected by internal processes, but also
influenced by external interactions with the surrounding environment
\citep{baughPrimerHierarchicalGalaxy2006, moGalaxyFormationEvolution2010}. From
this perspective, galaxies can be divided into central ones, which live in the
center of dark matter halos, and satellite ones, which are accreted by other
massive halo and move in the halo until they merge with the central galaxy in
the halo center \citep{wechslerConnectionGalaxiesTheir2018}. Various
astrophysical diagnostics are used to probe the impact of the environment on
galaxy evolution, including the gas-phase metallicity. At low $z$,
\citet{pasqualiGasphaseMetallicityCentral2012} found that satellite galaxies
have higher gas-phase metallicity than central ones with similar stellar mass
\citep[see also][]{pengDependenceGalaxyMassmetallicity2014,
donnanRoleCosmicWeb2022}. Similarly,
\citet{maierSlowthenrapidQuenchingTraced2019} found that star-forming cluster
galaxies have higher gas-phase metallicity than star-forming field galaxies,
even though no difference in the star formation rate (SFR) is seen, indicating
that gas-phase metallicity is more sensitive to the environment than SFR.

However, there is still a large diversity for the environmental dependence of
the gas-phase metallicity at high $z$. For example,
\citet{shimakawaEarlyPhaseEnvironmental2015} found that protocluster galaxies
at $z>2$ have higher gas-phase metallicity than those in the field by $\lesssim
0.15$ dex. \citet{calabroEnvironmentalDependenceStellar2022a} found that, at
$z\sim 3$, galaxies in over-dense regions have lower gas-phase metallicity than
those in the field by $\sim 0.1$ dex.
\citet{chartabMOSDEFSurveyEnvironmental2021} found that galaxies in the field
are more metal-rich than their counterparts in over-dense regions at $z\sim
2.3$, and the trend reverses at $z\sim 1.5$, where galaxies in over-dense
regions become more metal-rich. Recently,
\citet{wangMassMetallicityRelationCosmic2022} reported the result from the
first measurement of the mass-metallicity relation (MZR) at $z\sim 2.3$ in one
of the MAMMOTH protocluster region via grism spectroscopy
\citep[see][]{caiMAPPINGMOSTMASSIVE2016, caiMappingMostMassive2017,
    shiSpectroscopicConfirmationTwo2021, zhengMAMMOTHConfirmationTwo2021,
zhangSubmillimetreGalaxiesTwo2022}. They found that massive galaxies in
protoclusters have lower gas-phase metallicity than field galaxies, while, for
low-mass galaxies, those in protoclusters are more metal-rich.

On the theoretical side, current hydrodynamical simulations can provide some
insight into the underlying physics that drives the environmental dependence of
gas-phase metallicity \citep[e.g][]{schayeEAGLEProjectSimulating2015,
    pillepichSimulatingGalaxyFormation2018,
    daveSIMBACosmologicalSimulations2019,
rosas-guevaraRevealingPropertiesVoid2022, Metha.2023}.
\citet{baheOriginEnhancedMetallicity2017} studied the metallicity of
satellite galaxies in the EAGLE simulation, where they found that the metal
enrichment of satellite galaxies is driven by the suppression of
low-metallicity gas inflow and the stripping of low-metallicity gas from
the galaxy outskirt. \citet{guptaChemicalPreprocessingCluster2018} found
that galaxies are chemically enriched even when they have not been accreted
into clusters at $z < 1$.

In this paper, we use state-of-the-art hydrodynamical simulations to
investigate the environmental dependence of gas-phase metallicity from $z\sim
2.3$ to $z=0$. We will show how central and satellite galaxies play different
roles, as well as how the gas inflow and outflow processes regulate the
gas-phase metallicity. Finally, we will present the MZR difference in
protoclusters and the field, which agrees well with observational results in
\citet{wangMassMetallicityRelationCosmic2022}.

In \S\,\ref{sec:data}, we introduce the simulation data used in this paper. The
main results are presented in \S\,\ref{sec:results} and summarized in
\S\,\ref{sec:summary}. Throughout this paper, we converted all
cosmology-dependent quantities to a concordance $\Lambda$CDM cosmology with
$H_0=100h~\rm km/s/Mpc$, $h=0.7$, $\Omega_{\Lambda}=0.75$, and
$\Omega_{m}=0.25$.

\section{Data}%
\label{sec:data}

\begin{deluxetable*}{cccccccc}
    \tablenum{1}
    \tablecaption{
        Statistics of galaxies in different stellar mass and halo mass bins at
        three redshift snapshots.
    }
    \tablewidth{0pt}
    \tablehead{
        \colhead{} & \colhead{Stellar Mass} & \multicolumn{2}{c}{$z=0.0$} & \multicolumn{2}{c}{$z=1.0$} & \multicolumn{2}{c}{$z=2.3$}\\
        \colhead{} & \colhead{$\rm \log M_*/M_\odot$} & \colhead{$N$} & \colhead{$\log p$} & \colhead{$N$} & \colhead{$\log p$} & \colhead{$N$} & \colhead{$\log p$}
    }
    \startdata
     \multirow{10}{*}{Central}
     & 9.0 - 9.2   & 1171/0/0  & -0.0/-/-    & 1633/0/0  & -0.0/-/-     & 1409/1/0  & 0.0/-/-     \\
     & 9.2 - 9.4   & 1110/0/0  & -0.0/-/-    & 1344/0/0  & 0.0/-/-      & 1021/0/0  & 0.0/-/-     \\
     & 9.4 - 9.6   & 1022/0/0  & 0.0/-/-     & 1072/3/0  & 0.0/-/-      & 765/3/0   & 0.0/-/-     \\
     & 9.6 - 9.8   & 761/2/0   & 0.0/-/-     & 855/7/0   & 0.0/-/-      & 527/14/0  & -0.0/-5.9/- \\
     & 9.8 - 10.0  & 637/8/0   & 0.0/-/-     & 701/37/0  & -0.0/-8.2/-  & 386/42/0  & -0.1/-4.7/- \\
     & 10.0 - 10.2 & 513/56/0  & -0.3/-8.7/- & 485/128/0 & -1.8/-11.6/- & 298/138/0 & -2.9/-8.7/- \\
     & 10.2 - 10.4 & 284/216/0 & -3.6/-5.3/- & 256/288/1 & -7.6/-6.3/-  & 96/179/0  & -4.2/-1.7/- \\
     & 10.4 - 10.6 & 46/334/2  & -4.9/-0.1/- & 29/362/0  & -5.0/-0.0/-  & 5/189/0   & -/-0.0/-    \\
     & 10.6 - 10.8 & 1/248/4   & -/0.0/-     & 1/276/7   & -/-0.0/-     & 1/73/2    & -/0.0/-     \\
     & 10.8 - 11.0 & 0/156/29  & -/-0.3/-4.6 & 0/102/24  & -/-0.3/-3.8  & 0/37/5    & -/-0.0/-    \\
    \hline
     \multirow{10}{*}{Satellite}
      & 9.0 - 9.2   & 194/266/87 & -3.7/-0.8/-1.9 & 345/547/160 & -14.8/-0.9/-18.7 & 231/336/13 & -6.9/-3.3/-2.4 \\
      & 9.2 - 9.4   & 118/288/97 & -4.5/-0.1/-2.3 & 192/436/186 & -14.3/-0.3/-12.2 & 154/228/21 & -2.8/-1.0/-1.0 \\
      & 9.4 - 9.6   & 89/261/118 & -6.1/-0.0/-3.9 & 134/376/161 & -7.9/-1.1/-13.4  & 73/173/11  & -1.9/-0.2/-1.4 \\
      & 9.6 - 9.8   & 61/229/113 & -6.5/-0.7/-5.5 & 80/302/128  & -7.1/-0.8/-8.8   & 29/149/11  & -0.6/-0.0/-0.4 \\
      & 9.8 - 10.0  & 27/172/126 & -3.3/-1.6/-2.7 & 30/224/123  & -3.3/-1.9/-6.9   & 14/105/7   & -0.7/-0.0/-    \\
      & 10.0 - 10.2 & 13/153/111 & -1.9/-4.6/-4.5 & 16/199/105  & -1.2/-1.4/-3.8   & 7/76/12    & -/-0.0/-1.1    \\
      & 10.2 - 10.4 & 4/97/97    & -/-4.4/-2.7    & 5/148/91    & -/-0.7/-1.9      & 0/50/7     & -/-0.0/-       \\
      & 10.4 - 10.6 & 0/66/70    & -/-3.4/-0.5    & 0/86/78     & -/-1.4/-1.2      & 1/23/3     & -/-0.0/-       \\
      & 10.6 - 10.8 & 0/27/45    & -/-0.3/-0.0    & 0/28/65     & -/-0.6/-0.0      & 0/4/3      & -/-/-          \\
      & 10.8 - 11.0 & 0/2/31     & -/-/-0.0       & 0/3/24      & -/-/-0.0         & 0/3/0      & -/-/-          \\
    \enddata
    \tablecomments{
        Each cell contains three numbers, which are for galaxies with the host
        halo mass of $\rm\log (M_h/M_\odot)\in [11, 12),~[12, 13),~[13, 14)$,
        respectively.
        }
    \tablecomments{
        $N$ is the number of galaxies in each bin, and $\log p$ is the 10-based
        logarithm of the $p$ value from the KS test for the gas-phase
        metallicity residual between galaxies in each stellar mass bin and
        galaxies in each stellar mass and halo mass bin. $\log p<-1.3$
        indicates that the probability that the two samples in question come
        from the same distribution is smaller than 5\%.
    }
    \tablecomments{
        The $\log p$ values for bins with less than 10 galaxies are omitted.
        }
    \label{tab:statistics}
\end{deluxetable*}

We are using the EAGLE hydrodynamical simulation
\citep{schayeEAGLEProjectSimulating2015, crainEAGLESimulationsGalaxy2015,
mcalpineEagleSimulationsGalaxy2016, theeagleteamEAGLESimulationsGalaxy2017},
which consists of a suite of cosmological hydrodynamical simulations of
different box sizes, resolutions, and subgrid recipes. The analysis in this
paper is mainly based on the simulation labeled with \texttt{Ref-L100N1504}.
This specific simulation contains $1504^3$ dark matter particles with each
weighing $9.70\times 10^6M_{\odot}$ and $1504^3$ baryonic particles with each
weighing $1.81\times 10^6M_{\odot}$ in a (100cMpc)$^3$ cubic box.

Dark matter halos are identified by applying the Friends-of-friends (FoF)
method on all dark matter particles, and these halos are called main halos.
Subhalos are identified by applying the \texttt{SUBFIND} algorithm on all types
of particles \citep{springelPopulatingClusterGalaxies2001}. Then, for each
identified substructure, the baryonic part is defined as a galaxy and the dark
matter part is defined as a subhalo. In each FoF halo, the subhalo with the
lowest gravitational potential is defined as the central subhalo and the galaxy
in it is defined as the central galaxy, while others are satellite galaxies.
The subhalo merger trees are built with the SUBLINK algorithm
\citep{rodriguez-gomezMergerRateGalaxies2015}.

The mass for each main halo is defined as the total mass within a radius where
the mean density enclosed is 200 times the critical density. For each galaxy,
the stellar mass is defined as the total mass of star particles within 30 pkpc,
and SFR is defined as the sum of SFR for all star-forming gas particles within
the same aperture\footnote{ This stellar mass definition in the EAGLE
    simulation is recommended by \citet{schayeEAGLEProjectSimulating2015},
    where they emphasized that this aperture constraint can give results close
    to the Petrosian apertures that are often used for observation, which is
    used to tune the parameters in subgrid recipes in the EAGLE simulation.
    They also emphasized that this aperture effect is negligible for galaxies
    with $M_*<10^{11}M_{\odot}$, where Figure 9 in
    \citet{schayeEAGLEProjectSimulating2015} shows that the effective radius
    for galaxies with $M_*<10^{11}M_\odot$ is usually smaller than 10kpc.
    Meanwhile, we also found that our main results do not change if we adopted
    a different stellar mass definition, e.g. total stellar mass bound to the
subhalo.}. The specific SFR (SSFR) is defined as ${\rm SSFR}  = {\rm
SFR}/M_*$. The inter-stellar medium (ISM) mass for each galaxy is defined
as the total gas mass within the same aperture. The gas-phase metallicity
is quantified with the relative abundance of oxygen and hydrogen, $\rm\log
(O/H)$, for all star-forming gas particles. In this work, we only include
star-forming galaxies with non-zero SFR.

\section{Results}%
\label{sec:results}

In this section, we first presented the MZR measurements for simulated galaxies
in the EAGLE simulation, together with their environmental dependence, in
\S\,\ref{sub:the_environmental_dependence_of_mzr}. Then, we analyzed the
physical origin of the environmental dependence by studying the gas content,
halo mass, and black hole mass of these galaxies in
\S\,\ref{sub:physical_causes_of_the_environmental_dependence_of_mzr}. Finally,
we presented the MZR results at high $z$ in protocluster and field regions in
\S\,\ref{sub:mzr_in_high_z_protoclusters}.

\subsection{The environmental dependence of MZR}%
\label{sub:the_environmental_dependence_of_mzr}

\begin{figure*}
    \centering
    \includegraphics[width=0.9\linewidth]{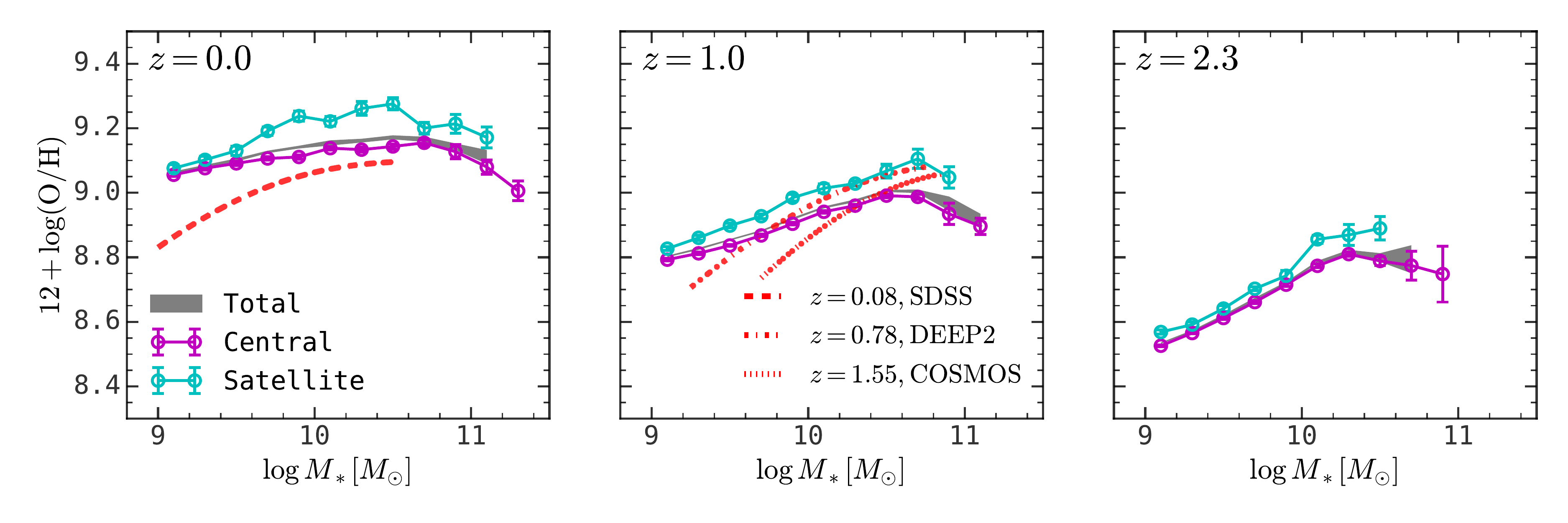}
    \caption{
        Gas phase metallicity as a function of stellar mass derived from the
        EAGLE simulations at $z=0$, $z=1$, and $z=2.3$. The shaded regions show
        the results for all of the galaxies, while the magenta and cyan symbols
        show the results for central and satellite galaxies, respectively.
        Error bars are estimated using the bootstrap method. The dashed,
        dash-dotted, and dotted red lines are fitting functions to
        observational results from \citet{zahidUniversalRelationGalactic2014}.
        Here one can see that the gas-phase metallicity monotonically increases
        towards low $z$ at fixed stellar mass. And the metallicity of satellite
        galaxies is higher than central galaxies by $\sim 0.1$ dex.
    }%
    \label{fig:figures/mzr_eagle}
\end{figure*}

\begin{figure*}
    \centering
    \includegraphics[width=0.9\linewidth]{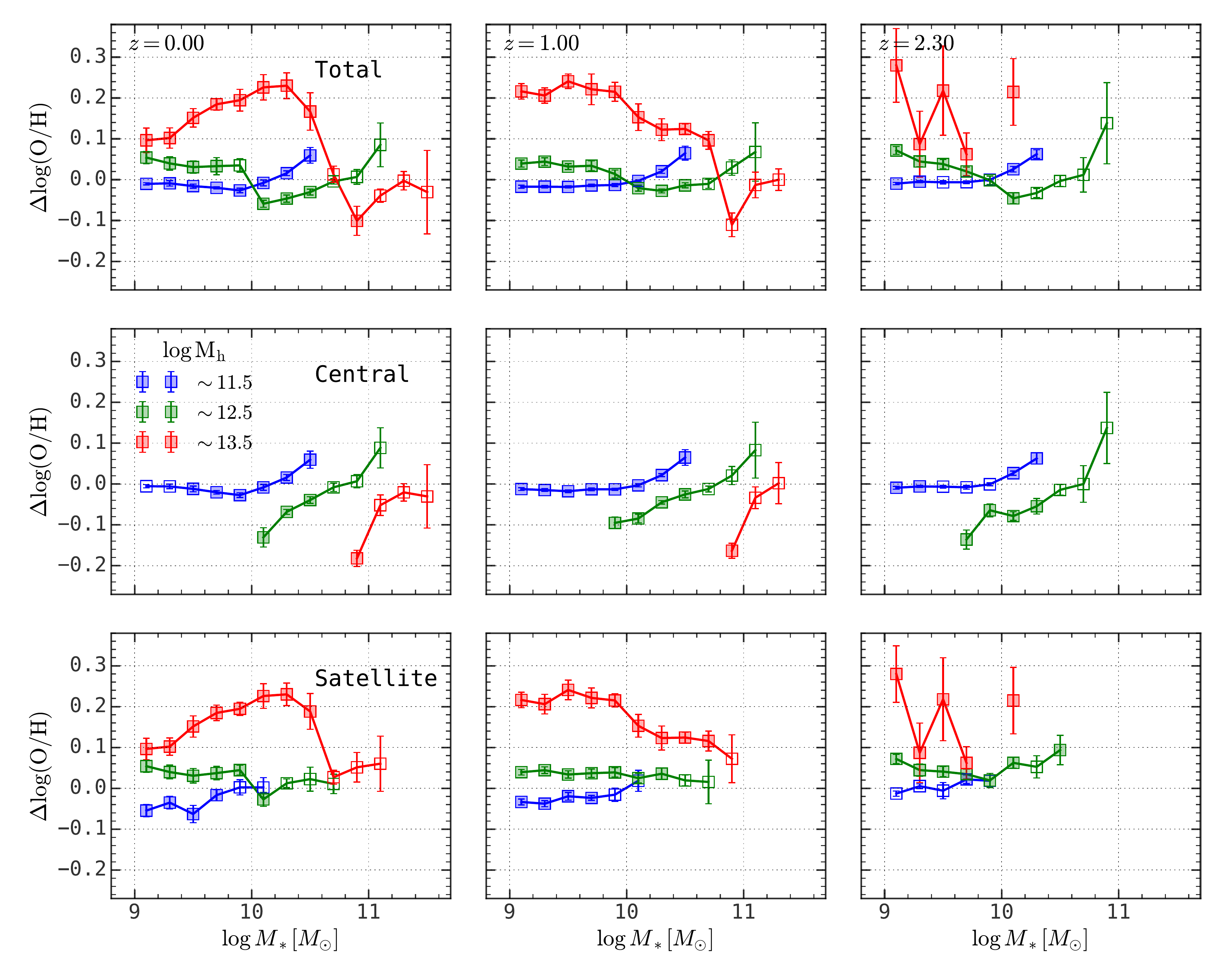}
    \caption{
        The median gas phase metallicity as a function of stellar mass for
        galaxies with different halo mass at $z=0$, $z=1$, and $z=2.3$. The
        number of galaxies in each bin is presented in
        Table~\ref{tab:statistics}. For clarity, we show the metallicity offset
        with respect to the average MZR in Figure~\ref{fig:figures/mzr_eagle}.
        The top panels show the results for all of the galaxies in three halo
        mass bins, while the middle and bottom panels show the results for
        central and satellite galaxies separately. Error bars show the standard
        deviation of the bootstrap sample. In each stellar mass bin, the
        difference in the gas-phase metallicity residual distribution for
        galaxies in each halo mass bin and those without halo constraint is
        quantified by a $p$ value from the KS test, as give in
        Table~\ref{tab:statistics}. Here we plot those points with $p \leq
        0.05$ using filled symbols and those with $p > 0.05$ using open
        symbols. Our results show that, with stellar mass fixed, central
        galaxies in massive halos are more metal-poor than those in
        less-massive halos, and the trend is reversed for satellite galaxies,
        where satellites in more massive halos are more metal-rich.
    }%
    \label{fig:figures/mzr_bin_z_eagle}
\end{figure*}

Figure~\ref{fig:figures/mzr_eagle} shows the gas-phase metallicity as a
function of stellar mass for galaxies at $z=0$, $z=1$, and $z=2.3$ in the EAGLE
simulation. The shaded regions are for all galaxies with non-zero SFR, and the
results of central and satellite galaxies are presented in magenta and cyan
symbols, respectively. The fitting results from
\citet{zahidUniversalRelationGalactic2014} are also shown in red curves. As one
can see, also noted by \citet{schayeEAGLEProjectSimulating2015}, the EAGLE
simulation over-predicts the gas-phase metallicity compared with observational
results at $z\sim 0$, especially for low-mass halos.
\citet{schayeEAGLEProjectSimulating2015} argued that observations tend to
under-estimate the gas-phase metallicity due to a non-negligible fraction of
metal condensed on dust grains
\citep[e.g.][]{dwekEvolutionElementalAbundances1998,
mattssonVarianceDusttometalsRatio2014}. Besides, gas-phase metallicity starts
to decline above $10^{10.5}M_\odot$ at $z\lesssim 1$ \citep[see
also][]{torreyEvolutionMassmetallicityRelation2019}, which is not seen in
observation. It is noteworthy that \citet{derossiGalaxyMetallicityScaling2017}
found that AGN feedback can effectively suppress the gas-phase metallicity for
these massive galaxies. Currently, there is no evidence showing that this
discrepancy depends on environment due to the lack of observational data at
$z\sim 1$. The next-generation surveys, like MOONS
\citep{maiolinoMOONRISEMainMOONS2020} and PFS
\citep{takadaExtragalacticScienceCosmology2014}, enable us to study the
environmental dependence of MZR \citep{wangIdentifyingGalaxyGroups2020,
wangFindingProtoclustersTrace2021} and compare them with simulations. At fixed
stellar mass, one can see that satellite galaxies are more chemically enriched
than central galaxies, consistent with previous observational and theoretical
results \citep[e.g.][]{ellisonMassmetallicityRelationGalaxy2009,
    daveGalaxyEvolutionCosmological2011, 
    pasqualiGasphaseMetallicityCentral2012,
    pengDependenceGalaxyMassmetallicity2014,
baheOriginEnhancedMetallicity2017}.

Figure~\ref{fig:figures/mzr_bin_z_eagle} shows the residual gas-phase
metallicity with respect to the median MZR in
Figure~\ref{fig:figures/mzr_eagle} (shaded region) for galaxies with different
stellar mass and host halo mass. At low-mass end, galaxies in massive halos are
more metal-rich than those in low-mass halos, and the trend is reversed for
massive galaxies, where those in massive halos are more metal-poor.
Figure~\ref{fig:figures/mzr_bin_z_eagle_finer_halo_mass_bin} presents the
results obtained in finer halo mass bins with a bin width of 0.3 dex, where one
can still see a similar trend as in Figure~\ref{fig:figures/mzr_bin_z_eagle},
showing that our results are not biased by the halo mass bin width adopted.

Table~\ref{tab:statistics} presents the number of galaxies in each stellar mass
and halo mass bin at three redshift snapshots for centrals and satellites,
respectively. For galaxies in each stellar mass bin, we also perform
Kolmogorov–Smirnov tests between those in different halo mass bins and all
galaxies in the whole host halo mass range, and the logarithmic $p$ values are
presented in Table~\ref{tab:statistics}. We note that a $\log p$ value smaller
than -1.3 means that the probability that two samples in question come from the
same distribution is smaller than 5\%, indicating that the difference between
these two samples are statistically significant. For example, for central
galaxies with $10^{10}M_\odot\leq M_* < 10^{10.2}M_\odot$ at $z=2.3$, we have
$\log p$ values of -2.9 and -8.7 for those with $10^{11}M_\odot\leq
M_h<10^{12}M_\odot$ and $10^{12}M_\odot\leq M_h<10^{13}M_\odot$, respectively,
indicating that the dependence of the gas-phase metallicity on host halo mass
for these central galaxies is not due to statistical fluctuation. Results in
Table~\ref{tab:statistics} confirms that the halo mass dependence shown in
Figure~\ref{fig:figures/mzr_bin_z_eagle} is statistically significant. For
clarity, we plot those points with $p \leq 0.05$ using filled symbols and those
with $p > 0.05$ using open symbols in Figure~\ref{fig:figures/mzr_bin_z_eagle}.

To understand the environmental dependence of gas-phase metallicity for
galaxies with different stellar mass, we present the gas-phase metallicity
residuals for central and satellite galaxies separately on the middle and
bottom panels of Figure~\ref{fig:figures/mzr_bin_z_eagle}. For central
galaxies, those in massive halos are more metal-rich than their counterparts in
low-mass halos. On the contrary, satellite galaxies in more massive halos are
more metal-rich than those in low-mass halos. Combined with that the satellite
fraction monotonically decreases with increasing stellar mass
\citep[e.g.][]{zhengTheoreticalModelsHalo2005, yangGalaxyGroupsSDSS2008}, the
environmental dependence of gas-phase metallicity shown on top panels can be
explained by the relative abundance of central and satellite galaxies at
different stellar mass bins, where central galaxies dominate the massive end
and satellite galaxies dominate the low-mass end.

It is noteworthy that the IllustrisTNG simulation can also produce similar
environmental dependence of gas-phase metallicity for central and satellite
galaxies, as shown in
Appendix~\ref{sec:the_environmental_dependence_of_mzr_in_illustristng} and
Figure~\ref{fig:figures/mzr_bin_z_tng100}, despite the difference in their
subgrid recipes.

\subsection{Physical causes of the environmental dependence of MZR}%
\label{sub:physical_causes_of_the_environmental_dependence_of_mzr}

The gas-phase metallicity is regulated by a variety of processes \citep[see
e.g.][]{finlatorOriginGalaxyMassmetallicity2008,
    lillyGASREGULATIONGALAXIES2013, pipinoRelationSpecificStar2014,
    dekelAnalyticSolutionMinimal2014, feldmannEquilibriumViewDust2015,
maiolinoReMetallicaCosmic2019, linConstraintsGalacticOutflows2023}, which can
be grouped into three categories. The first one is star formation. This process
consumes cold gas to form new stars, which can reduce the cold gas amount.
Moreover, the stellar wind and the supernovae feedback can enrich the gas-phase
metal content by returning metal into the ISM. The overall effect of the star
formation activity is to enrich the metal content in ISM.

The second one is gas inflow processes through either gas accretion or galaxy
merger. The impact on the global gas-phase metallicity depends on the relative
metallicity of the accreted gas and the existing ISM. Previous studies found
that usually accreted gas is more chemically pristine than the ISM
\citep[e.g.][]{wrightRevealingPhysicalProperties2021}, so that the gas inflow
can dilute the ISM content and reduce the gas-phase metallicity
\citep{ceverinoGasInflowMetallicity2016}.

The last one is gas outflow driven by various processes
\citep[e.g.][]{finlatorOriginGalaxyMassmetallicity2008,
    boucheMissingMetalsProblem2007, daveGalaxyEvolutionCosmological2011,
feldmannEquilibriumViewDust2015, trusslerBothStarvationOutflows2020}. First of
all, gas outflows can reduce the ISM amount so that the gas-phase metallicity
becomes easier to be altered by other processes. Then, some outflow processes
prefer to eject gas whose metallicity deviates from the globally averaged value
so that the global metallicity can be increased or decreased, depending on the
sign of the deviation. For example, the AGN feedback process prefer to
accelerate/heat gas in the galaxy center so that these gas are preferentially
ejected. On the contrary, the ram-pressure stripping effect prefers to affect
gas on the galaxy outskirt. If the galaxy of interest exhibit a non-zero
metallicity gradient \citep[e.g.][]{wangGrismLensamplifiedSurvey2017,
    wangDiscoveryStronglyInverted2019, wangCensusSubkiloparsecResolution2020,
    collacchioniEffectGasAccretion2020, hemlerGasphaseMetallicityGradients2021,
    tisseraEvolutionOxygenAbundance2022, wangEarlyResultsGLASSJWST2022,
liFirstCensusGasphase2022}, the above two outflow processes can also change the
global metallicity \citep{vanloonExplainingScatterGalaxy2021}. It is noteworthy
that the differential wind assumption, where metals are more efficiently
ejected than other elements through outflow, is adopted by many analytical
models to explain observational results
\citep[e.g.][]{recchiEffectDifferentialGalactic2008,
dayalPhysicsFundamentalMetallicity2013}.

Next, we will explore the physical causes for the environmental dependence of
gas-phase metallicity from these three aspects for central and satellite
galaxies, separately.

\subsubsection{Central galaxies}%
\label{ssub:central_galaxies}

\begin{figure*}
    \centering
    \includegraphics[width=0.9\linewidth]{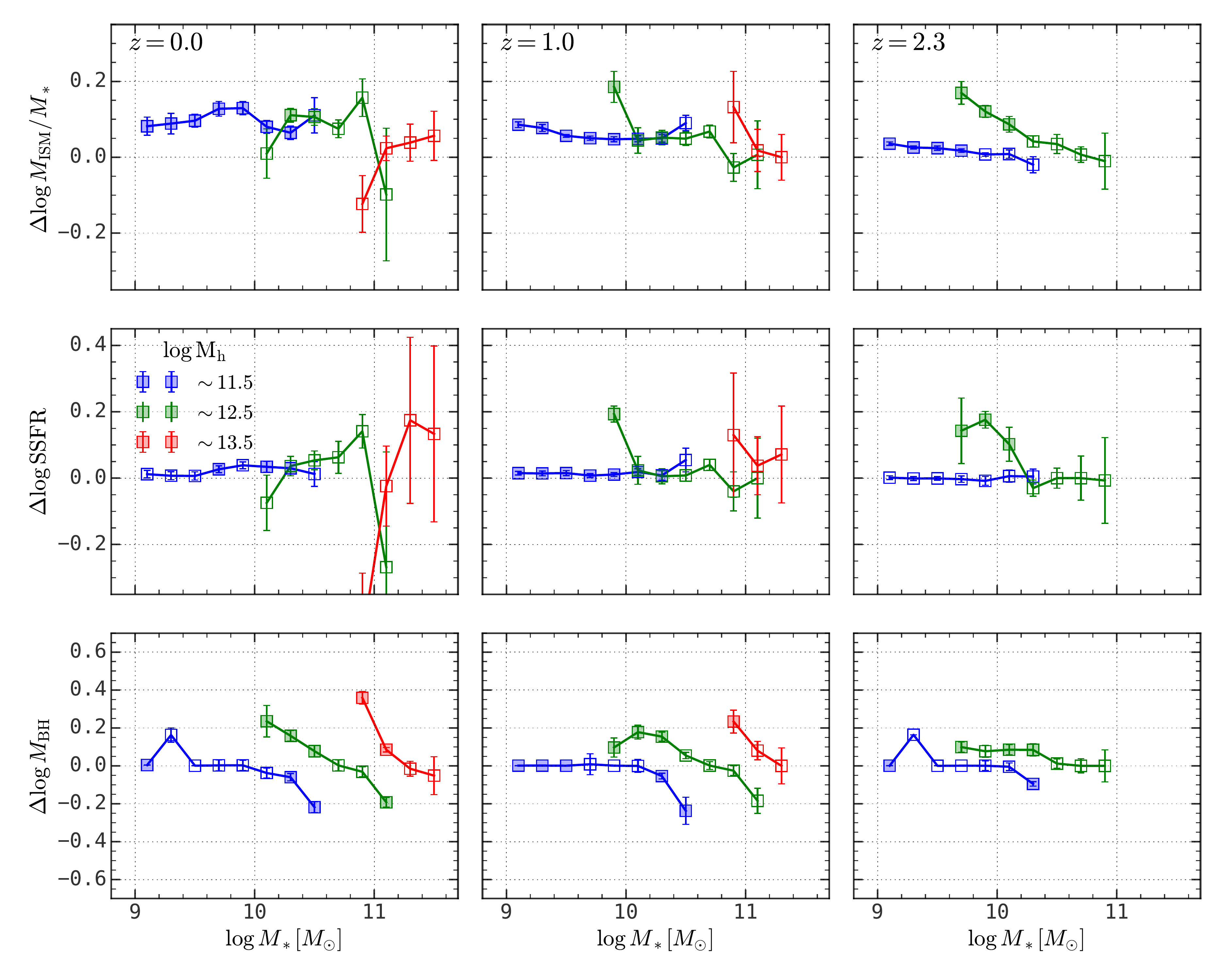}
    \caption{
        Properties of central galaxies as a function of stellar mass in three
        halo mass bins at $z=0$, $z=1$, and $z=2.3$ from the EAGLE simulation
        ({\bf Top panels:} The median ISM mass to stellar mass ratio; {\bf
        Middle panels:} The median specific star formation rate; {\bf Bottom
        panels:} The median black hole mass). All three properties are
        presented as the residual with respect to the median value in each
        stellar mass bin for clear demonstration. Error bars are calculated
        using the bootstrap method. The difference between filled and open
        symbols are similar to Figure~\ref{fig:figures/mzr_bin_z_eagle}, except
        here the $p$-values are calculated for the residual ISM mass fraction,
        SSFR, and black hole mass. At high $z$, central galaxies in more
        massive halo possess more ISM, while, at low $z$, central galaxies in
        more massive halo harbor more massive black holes.
    }%
    \label{fig:figures/combined_results_cen_eagle}
\end{figure*}

Figure~\ref{fig:figures/mzr_bin_z_eagle} shows that, at $z\sim 2.3$, central
galaxies in massive halos are more metal-poor than those in low-massive halos.
The gas-phase metallicity difference is about 0.1 dex for central galaxies in
$\sim 10^{11.5}M_{\odot}$ and $\sim 10^{12.5}M_{\odot}$ halos.
Figure~\ref{fig:figures/combined_results_cen_eagle} shows the difference of
other properties as a function of stellar mass for central galaxies in
different halo mass bins, where one can see that central galaxies in massive
halos have more abundant ISM content, which can dilute the metallicity if the
excess fraction of ISM is metal-poor. Actually,
\citet{mitchellGalacticInflowWind2020} studied the gas accretion onto central
galaxies in the EAGLE simulation, where they found that the gas accretion rate
onto central galaxies at $z\sim 2$ is almost proportional to their host halo
mass, and $\gtrsim 50\%$ of these accreted gas is pristine \citep[see
also][]{keresHowGalaxiesGet2005, dekelColdStreamsEarly2009,
keresGalaxiesSimulatedLCDM2009, zhangscience2023}. Consequently, at given
stellar mass, central galaxies living in massive halos can obtain more
low-metallicity gas through accretion, which can effectively dilute the
gas-phase metallicity of these galaxies. Besides,
Figure~\ref{fig:figures/combined_results_cen_eagle} also shows that central
galaxies in more massive halos are more actively forming new stars than those
in low-mass halos, which is expected from FMR
\citep[e.g.][]{ellisonCluesOriginMassMetallicity2008,
mannucciFundamentalRelationMass2010}. Even though the excess star formation
activities can chemically enrich the ISM, it still cannot compensate the
dilution effect brought by the gas inflow. Finally, we find that massive halos
prefer to host more massive black holes than those low-mass halos by $\sim 0.1$
dex. In order to figure out whether central super-massive black holes and their
feedback effects play a role here, we compare the environmental dependence for
simulations with and without AGN feedback in
Figure~\ref{fig:figures/noagn_eagle_mzr}, where one can see that the dependence
of MZR on halo mass is still present when the AGN feedback is disabled (dashed
lines) at $z\sim 2.3$. This indicates that AGN feedback effects only play a
minor role here.

Figure~\ref{fig:figures/mzr_bin_z_eagle} shows that central galaxies in massive
halos at $z=0$ have lower gas-phase metallicity than their counterparts in
low-mass halos by 0.1-0.2 dex. However, as shown in
Figure~\ref{fig:figures/combined_results_cen_eagle}, there is nearly no
difference in their ISM content, so the dilution effect discussed above does
not apply here. Neither do we see any obvious difference in their SSFR.
Finally, there is a strong dependence of black hole mass on the host halo mass,
where central galaxies in massive halos prefer to harbor more massive black
holes than their counterparts in low-mass halos. In the EAGLE simulation, due
to the simple one-mode AGN feedback recipe adopted, the feedback energy is
proportional to the mass of black holes, so that central galaxies in massive
halos, which possess more massive black holes, are able to trigger more
powerful AGN feedback and expel more gas out of the halo. This scenario is
supported by \citet{mitchellGalacticInflowWind2020}, where they found that the
AGN feedback is responsible for the majority of gas outflow on both galaxy and
halo scales for halos with $M_h\geq 10^{12}M_{\odot}$. The impact of gas
outflow induced by AGN feedback on the metallicity of central galaxies is
two-fold. First, the EAGLE simulation predict negative gradients in the spatial
distribution of gas-phase metallicity for central galaxies
\citep[see][]{baheOriginEnhancedMetallicity2017,
tisseraOxygenAbundanceGradients2019, tisseraEvolutionOxygenAbundance2022}, and
the AGN feedback preferentially heat the gas in the galaxy center, so that the
AGN feedback process is able to expel these metal-rich gas in the galaxy center
and decrease the global metallicity. Second, since we did not see any
difference in the ISM content for central galaxies in different mass of halos,
the ISM deficiency caused by AGN feedback is compensated by the accretion of
gas, which is more metal-poor than the existing gas in the galaxy and can
effectively dilute the metal content in the ISM.

\citet{yangImpactGasAccretion2022} studied the relation between the gas-phase
metallicity and the halo mass for central galaxies at fixed stellar mass using
clustering strength to indicate halo mass. For massive galaxies above
$10^{10}M_\odot$, they detect no clustering difference due to poor statistics,
while both EAGLE and IllustrisTNG predict more massive halos prefer to host
metal-poor central galaxies at fixed stellar mass. More observational data is
required to test the prediction of these hydrodynamical simulations.

At $z=1$, one can simultaneously see the excess of the ISM content and more
massive black holes for central galaxies in massive halos, showing that both
the accretion of low-metallicity gas and the AGN feedback play a role in
reducing their global metallicity.

In Appendix~\ref{sec:mzr_for_central_galaxies_without_agn_feedback} and
Figure~\ref{fig:figures/noagn_eagle_mzr}, we present the gas-phase metallicity
difference of central galaxies with different halo mass in a simulation where
the AGN feedback is disabled. Compared with the reference simulation, the
no-AGN case exhibits less environmental dependence of the gas phase metallicity
for central galaxies at $z=0$, while the dependence at high $z$ is unchanged.
This result indicates that AGN feedback does play a role in amplifying the MZR
difference for central galaxies with different halo mass at low $z$. And its
impact becomes negligible at high $z$.

\subsubsection{Satellite galaxies}%
\label{ssub:satellite_galaxies}

\begin{figure*}
    \centering
    \includegraphics[width=0.9\linewidth]{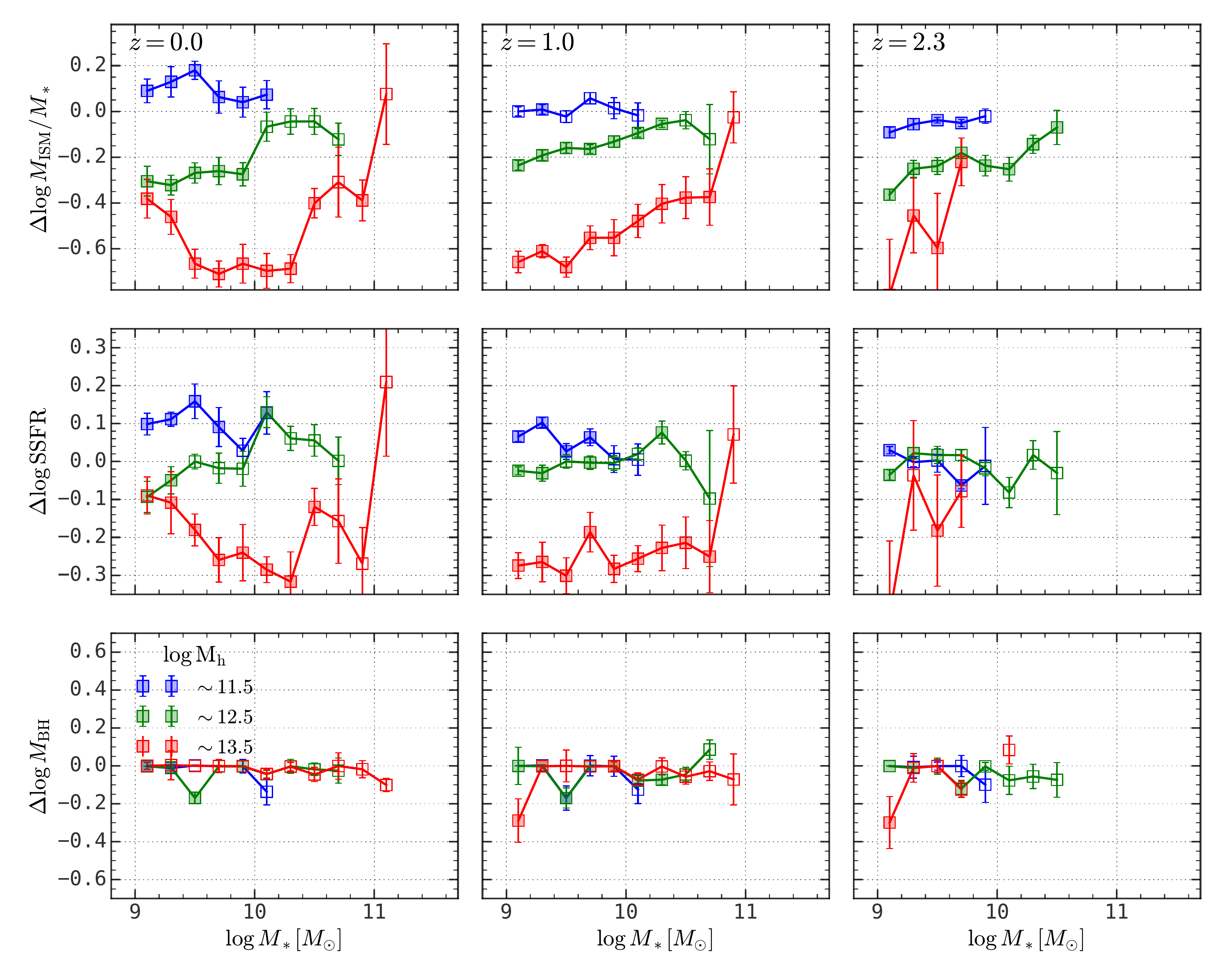}
    \caption{
        Similar to Figure~\ref{fig:figures/combined_results_cen_eagle}, except
        for satellite galaxies. Satellite galaxies in more massive halos have
        less ISM due to strangulation and stripping, and consequently, their
        SSFR is decreased. Meanwhile, there is no halo mass dependence of black
        hole mass for satellite galaxies.
    }%
    \label{fig:figures/combined_results_sat_eagle}
\end{figure*}

Compared with central galaxies, satellite galaxies are also subject to various
environmental effects. First of all, when a galaxy falls into other massive
halos and becomes a satellite galaxy, its accretion of dark matter and baryons
is reduced, or even terminated
\citep[e.g.][]{behrooziMergersMassAccretion2014}, and its ambient gas, i.e. the
circum-galactic medium (CGM), can be stripped, so that its ISM replenishment is
blocked. Without the dilution from the accreting gas, the gas-phase metallicity
increases persistently during the star-forming phase, until the cold gas
content is consumed and the galaxy is quenched. This effect is called
strangulation \citep[e.g.][]{larsonEvolutionDiskGalaxies1980,
pengStrangulationPrimaryMechanism2015}. In addition, satellite galaxies also
suffer from the ram-pressure stripping effect which can strip the ISM content
out of these satellite galaxies when they are moving in the hot
intra-group/cluster medium of host halos
\citep[e.g.][]{gunnInfallMatterClusters1972, vulcaniComparingRelationStar2010,
vulcaniGrismLensAmplifiedSurvey2016, poggiantiGASPGasStripping2017}. This
effect preferentially affect the gas on the outskirt of satellite galaxies
since it is relatively loosely bound compared with those in the galaxy center
\citep[e.g.][]{boselliRamPressureStripping2022}.
\citet{baheOriginEnhancedMetallicity2017} studied the environmental dependence
of the gas-phase metallicity for satellite galaxies at $z=0$ in the EAGLE
simulation, and they identified that both the stripping of low-metallicity gas
on the satellite outskirt and the lack of low-metallicity gas replenishment
into the galaxy center through accretion are the key drivers of the metallicity
enhancement in satellite galaxies.

Figure~\ref{fig:figures/mzr_bin_z_eagle} shows that satellite galaxies in
massive halos are more metal-rich than those in low-mass halos by $\lesssim
0.2$ dex, indicating that those environmental effects are already operating on
satellite galaxies from $z\sim 2.3$.
Figure~\ref{fig:figures/combined_results_sat_eagle} shows the ISM content, SSFR
and black hole mass for satellite galaxies as a function of stellar mass in
different mass of host halos. The ISM content is mostly stripped for satellite
galaxies in massive halos, and SSFR is also reduced. These results are
consistent with the two drivers identified in
\citet{baheOriginEnhancedMetallicity2017}, which can reduce the ISM content and
diminish SSFR simultaneously. Finally, the black hole mass for satellite
galaxies has no dependence on the host halo mass when the stellar mass is
fixed, as shown in the bottom panels of
Figure~\ref{fig:figures/combined_results_sat_eagle}, indicating that AGN
feedback is not important in setting the environmental dependence of MZR for
satellite galaxies.

\subsection{MZR in high-$z$ protoclusters}%
\label{sub:mzr_in_high_z_protoclusters}

\begin{figure*}
    \centering
    \includegraphics[width=0.9\linewidth]{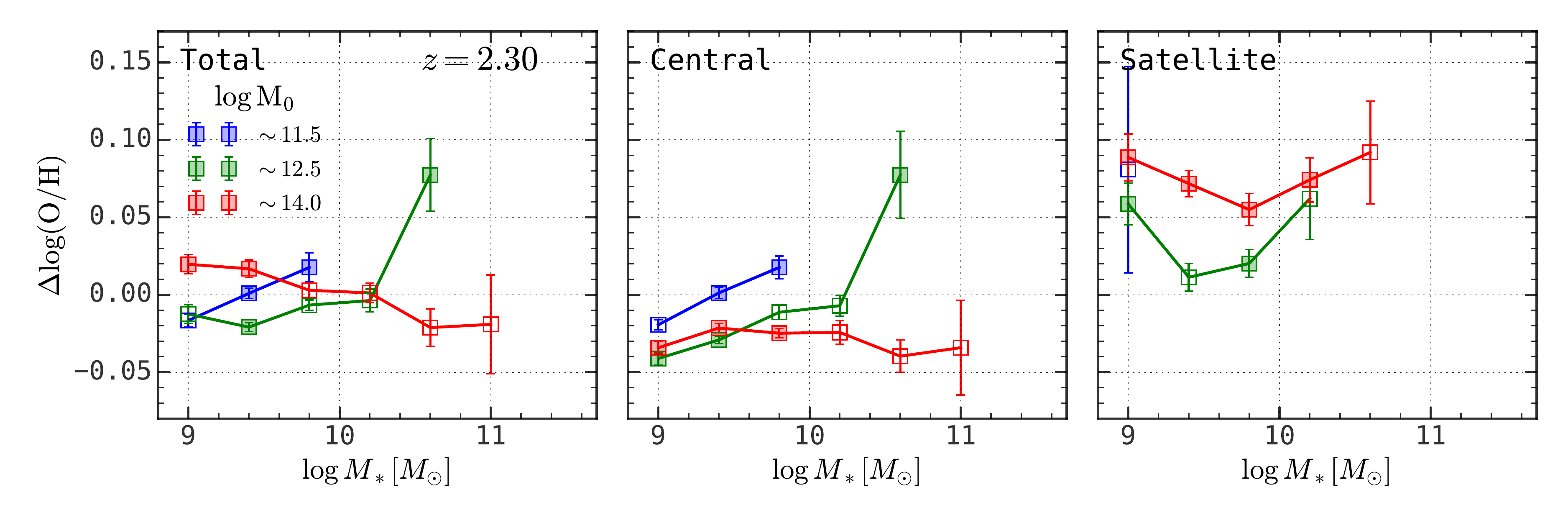}
    \caption{
        The residual MZR for galaxies at $z\sim 2.3$ with descendant halo mass,
        i.e. $\log (M_0/M_{\odot})$, in bins of [11, 12), [12, 13), and [13,
        15), and they are shown in blue, green, and red colors, respectively.
        The difference between filled and open symbols are similar to
        Figure~\ref{fig:figures/mzr_bin_z_eagle}, except here we are binning in
        descendant halo mass instead of halo mass. Central galaxies in
        protoclusters (high $M_0$) have lower metallicity and dominate the
        high-$M_*$ end, while satellite galaxies in protoclusters have higher
        metallicity and dominate the low-$M_*$ end.
    }%
    \label{fig:figures/mzr_pc_eagle}
\end{figure*}

\begin{figure}
    \centering
    \includegraphics[width=0.9\linewidth]{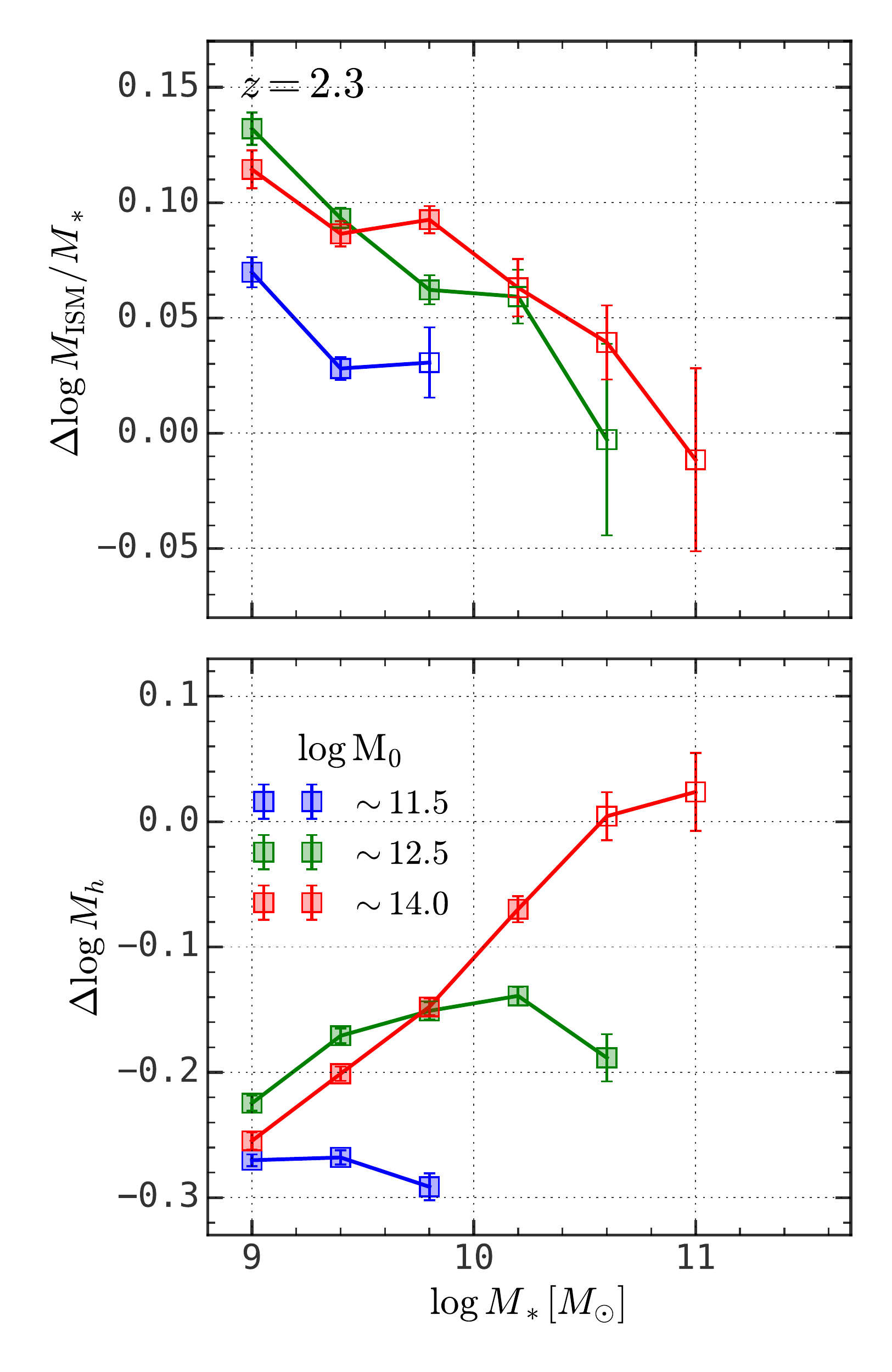}
    \caption{
        The residual ISM mass fraction ({\bf upper panel}) and host halo mass
        ({\bf lower panel}) for central galaxies at $z\sim 2.3$ with descendant
        halo mass, i.e. $\log (M_0/M_{\odot})$, in bins of [11, 12), [12, 13),
        and [13, 15), and they are shown in blue, green, and red colors,
        respectively. The difference between filled and open symbols are
        similar to Figure~\ref{fig:figures/combined_results_cen_eagle}, except
        here we are binning in descendant halo mass instead of halo mass.
        Central galaxies in protoclusters (high $M_0$) tend to live in more
        massive halos and possess more abundant ISM than those in the field
        (low $M_0$).
    }%
    \label{fig:figures/cen_prop_in_pc_eagle}
\end{figure}

\begin{figure}
    \centering
    \includegraphics[width=0.9\linewidth]{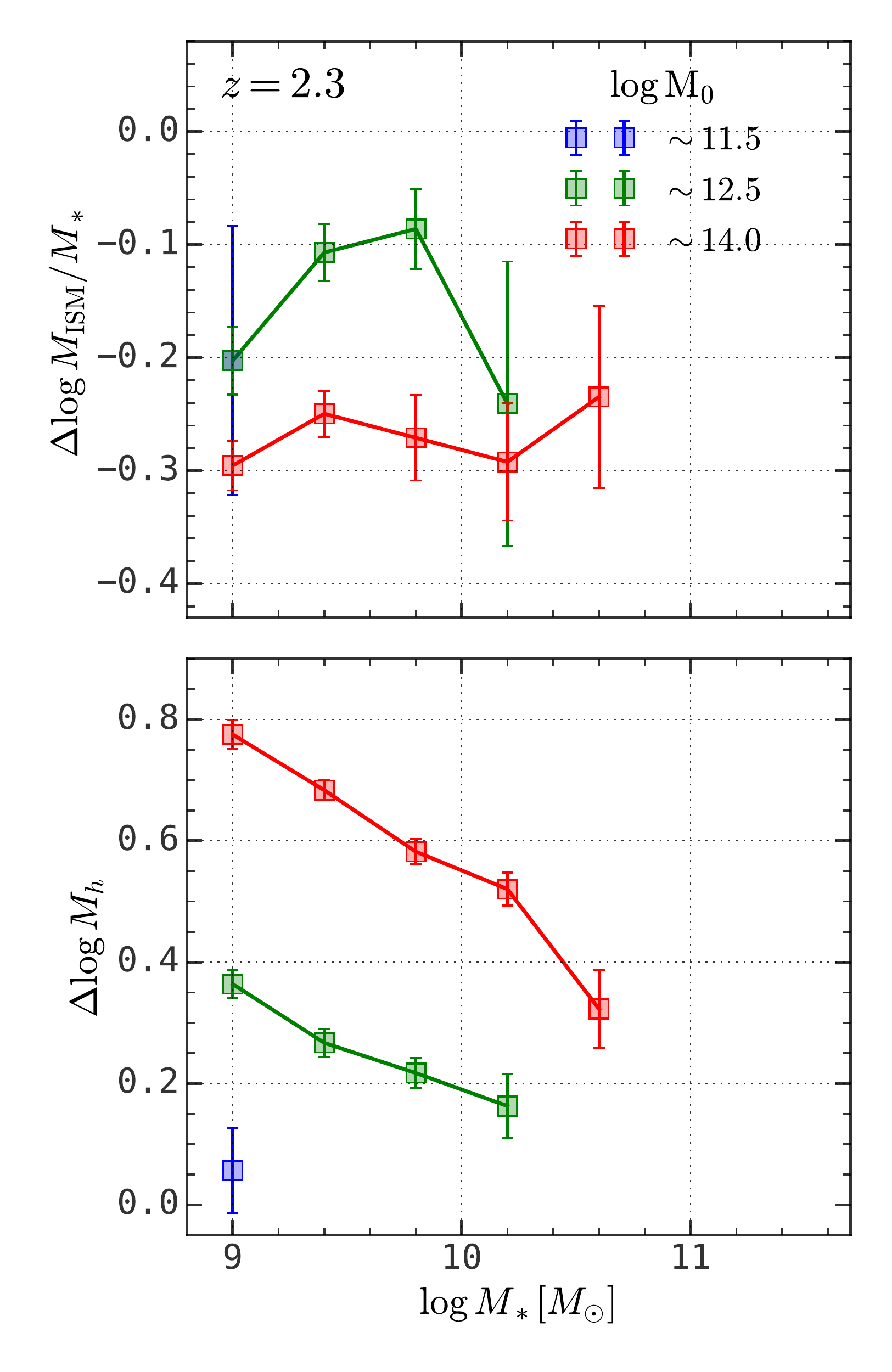}
    \caption{
        Similar to Figure~\ref{fig:figures/cen_prop_in_pc_eagle}, except for
        satellite galaxies. Satellite galaxies in protoclusters (high $M_0$)
        prefer to live in more massive halos and have lower ISM content than
        those in the field (low $M_0$).
    }%
    \label{fig:figures/sat_prop_in_pc_eagle}
\end{figure}

\citet{wangMassMetallicityRelationCosmic2022} found that massive galaxies in
protocluster region are more metal-poor than those in the field region. And the
trend is reversed for low-mass galaxies, where those in protoclusters are more
metal-rich. Here we want to see whether the EAGLE simulation can produce
similar results. To this end, we trace each galaxy at $z\sim 2.3$ to $z=0$ and
define the host halo mass there as the descendant halo mass, which is denoted
as $M_0$. Then, we can study the dependence of gas-phase metallicity on $M_0$.

Figure~\ref{fig:figures/mzr_pc_eagle} shows the gas-phase metallicity as a
function of stellar mass for galaxies at $z\sim 2.3$ in different descendant
halo mass bins. As one can see here, EAGLE predicts a similar environmental
dependence of MZR to the result reported in
\citet{wangMassMetallicityRelationCosmic2022}, where massive galaxies are more
metal-rich in the field (low $M_0$) and low-mass galaxies are more metal-rich
in protoclusters (high $M_0$). By separating the contribution from central and
satellite galaxies, one can see that central galaxies with high $M_0$ are more
metal-poor, which dominates the massive end, while satellite galaxies with high
$M_0$ are more metal-rich, which dominates the low-mass end.

Figure~\ref{fig:figures/cen_prop_in_pc_eagle} shows that central galaxies end
up in massive halos (high $M_0$) prefer to live in massive halos at high $z$
and possess more abundant ISM. This is consistent with the picture presented in
\S\,\ref{ssub:central_galaxies}, where galaxies in more massive halos can
accumulate more abundant gas through cold-mode accretion which can effectively
dilute the gas-phase metallicity.

Figure~\ref{fig:figures/sat_prop_in_pc_eagle} shows that satellite galaxies end
up in massive halos (high $M_0$) prefer to be satellite galaxies of massive
halos at high $z$, and, consequently, their ISM is reduced more by various
environmental effects. These results are consistent with our analysis in
\S\,\ref{ssub:satellite_galaxies}.

\section{Summary}%
\label{sec:summary}

Gas-phase metallicity is an important probe for a variety of astrophysical
processes in galaxy formation and evolution. Here we use the state-of-the-art
hydrodynamical simulations to investigate the environmental dependence of
gas-phase metallicity for central and satellite galaxies separately, as well as
their physical origins. Our main results are summarized as follows.
\begin{enumerate}
    \item From $z\sim 0$ to $z\sim 2.3$, massive galaxies in low-mass halos are
        more metal-rich than their counterparts in massive halos. For low-mass
        galaxies, the trend reverses, where those in more massive halos are
        more metal-rich. By separating the contribution from central and
        satellite galaxies, we find that the environmental dependence at the
        massive end is dominated by central galaxies, while satellite galaxies
        dominate the trend at the low-mass end (see
        Figure~\ref{fig:figures/mzr_bin_z_eagle} and
        Figure~\ref{fig:figures/mzr_bin_z_tng100}).

    \item  We found that, at high $z$, central galaxies in massive halos tend
        to have more ISM than those in low-mass ones with stellar mass fixed.
        The difference in the ISM fraction is presumably responsible for the
        halo mass dependence of central galaxy metallicity. At low $z$, the AGN
        feedback process also helps to decrease the metallicity of central
        galaxies in massive halos by ejecting metal-rich gas (see
        Figure~\ref{fig:figures/combined_results_cen_eagle} and
        Figure~\ref{fig:figures/noagn_eagle_mzr}).

    \item For satellite galaxies, the strangulation effect and the striping of
        ISM is responsible for the metallicity difference in satellite galaxies
        in different host halo mass bins, which is consistent with previous
        studies. Moreover, we did not see any difference in their black hole
        mass, indicating that the AGN feedback effect is not important here
        (see Figure~\ref{fig:figures/combined_results_sat_eagle}).

    \item We also studied the dependence of gas-phase metallicity for high-$z$
        galaxies on their descendant halo mass, $M_0$. We found that massive
        galaxies in protoclusters (high $M_0$) are more metal-poor than those
        in the field (low $M_0$). And low-mass galaxies exhibit a reversed
        trend, where those in protoclusters are more metal-rich. This result
        qualitatively agrees with the observational results in
        \citet{wangMassMetallicityRelationCosmic2022} (see
        Figure~\ref{fig:figures/mzr_pc_eagle} and
        Figure~\ref{fig:figures/mzr_pc_tng100}).

    \item The $M_0$-dependence of gas-phase metallicity for high-$z$ galaxies
        is a combined effect of central and satellite galaxies. The massive and
        low-mass ends are dominated by central and satellite galaxies,
        respectively. Meanwhile, central galaxies in protoclusters prefer to
        live in more massive halos and possess more ISM than their counterparts
        in the field. Satellite galaxies in protoclusters prefer to live in
        more massive halos so that they suffer more from environmental effects
        (see Figure~\ref{fig:figures/cen_prop_in_pc_eagle} and
        Figure~\ref{fig:figures/sat_prop_in_pc_eagle}).

\end{enumerate}

In this work, we investigated the environmental dependence of gas-phase
metallicity on the host halo mass for central and satellite galaxies using the
hydrodynamical simulations of EAGLE and IllustrisTNG. We also found that the
environmental dependence of gas-phase metallicity at high $z$ can be
qualitatively explain by both simulations.

Both simulations also predict that the scatter in the stellar mass-halo mass
relation correlates with the gas-phase metallicity, in the sense that galaxies
in more massive halos are more metal-poor \citep[see][for correlation with
other galaxy properties]{wangLateformedHaloesPrefer2023,
cuiOriginGalaxyColour2021, zhangMassiveStarformingGalaxies2022}. This
prediction can be tested in our local Universe and high $z$ using the galaxy
group catalog \citep{yangGalaxyGroupsSDSS2008, wangIdentifyingGalaxyGroups2020,
yangExtendedHalobasedGroup2021, liGroupsProtoclusterCandidates2022}. Besides,
the observational measurements of the gas-phase metallicity difference between
protocluster and field galaxies can be further improved with the coming
high-$z$ spectroscopic galaxy surveys
\citep{takadaExtragalacticScienceCosmology2014, maiolinoMOONRISEMainMOONS2020,
wangFindingProtoclustersTrace2021}.

Finally, it is noteworthy that the currently available sample of metallicity
measurements in over-dense regions still suffers from the small number
statistics which precludes a more quantitative comparison with our theoretical
investigation presented here. With more observational results to be obtained
from the ongoing MAMMOTH-Grism survey (HST-GO-16276, P. I. Wang), we can derive
a more robust scaling relation of metallicity offset versus overdensity, in
order to compare with our theoretical models in a more meaningful way.

\section*{Acknowledgements}

The authors thank the anonymous referee for their helpful comments that
improved the quality of the manuscript. The authors acknowledge the Tsinghua
Astrophysics High-Performance Computing platform at Tsinghua University for
providing computational and data storage resources that have contributed to the
research results reported within this paper. This work is supported by the
National Science Foundation of China (NSFC) Grant No. 12125301, 12192220,
12192222, and the science research grants from the China Manned Space Project
with NO. CMS-CSST-2021- A07. XW is supported by CAS Project for Young
Scientists in Basic Research, Grant No. YSBR-062.

\section*{Data availability}

The data underlying this article will be shared on reasonable request to the
corresponding author. The computation in this work is supported by the HPC
toolkit \specialname[HIPP] \citep{hipp}.

\bibliographystyle{aasjournal}
\bibliography{bibtex.bib}

\appendix

\section{The environmental dependence of MZR in finer halo mass bins}%
\label{sec:the_environmental_dependence_of_mzr_in_finer_halo_mass_bins}

\begin{figure*}
    \centering
    \includegraphics[width=0.9\linewidth]{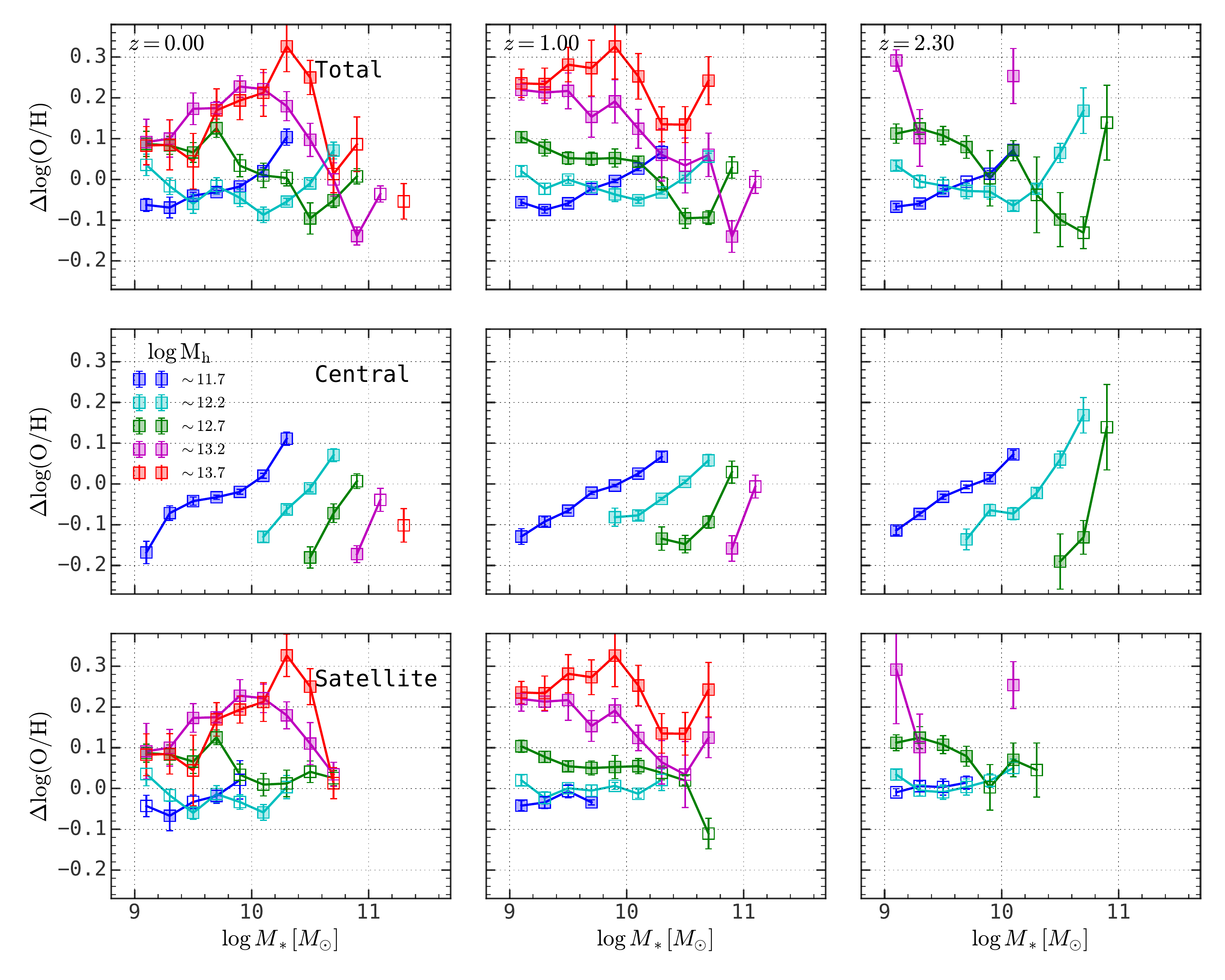}
    \caption{
        Similar to Figure~\ref{fig:figures/mzr_bin_z_eagle}, except with a halo
        mass bin width of 0.3 dex.
    }%
    \label{fig:figures/mzr_bin_z_eagle_finer_halo_mass_bin}
\end{figure*}

Figure~\ref{fig:figures/mzr_bin_z_eagle_finer_halo_mass_bin} shows the residual
MZR in different halo mass bins at three redshift snapshots for total, central,
and satellite galaxies, respectively. Here we adopted finer halo mass bins with
a width of 0.3 dex, where one can still see the trends shown in
Figure~\ref{fig:figures/mzr_bin_z_eagle}.

\section{The environmental dependence of MZR in IllustrisTNG}%
\label{sec:the_environmental_dependence_of_mzr_in_illustristng}

\begin{figure*}
    \centering
    \includegraphics[width=0.9\linewidth]{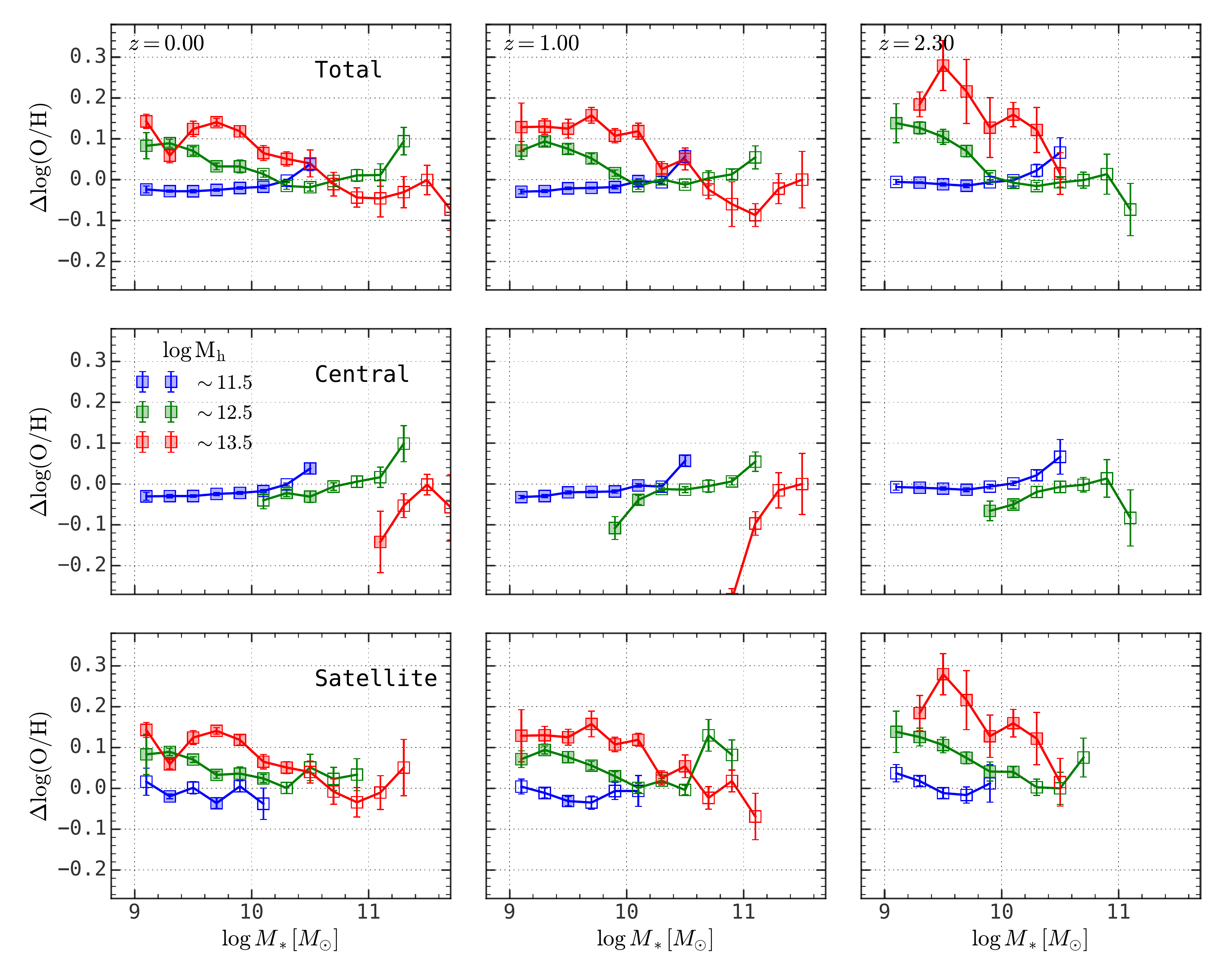}
    \caption{
        Similar to Figure~\ref{fig:figures/mzr_bin_z_eagle}, except for the
        TNG100 simulation.
    }%
    \label{fig:figures/mzr_bin_z_tng100}
\end{figure*}

\begin{figure*}
    \centering
    \includegraphics[width=0.9\linewidth]{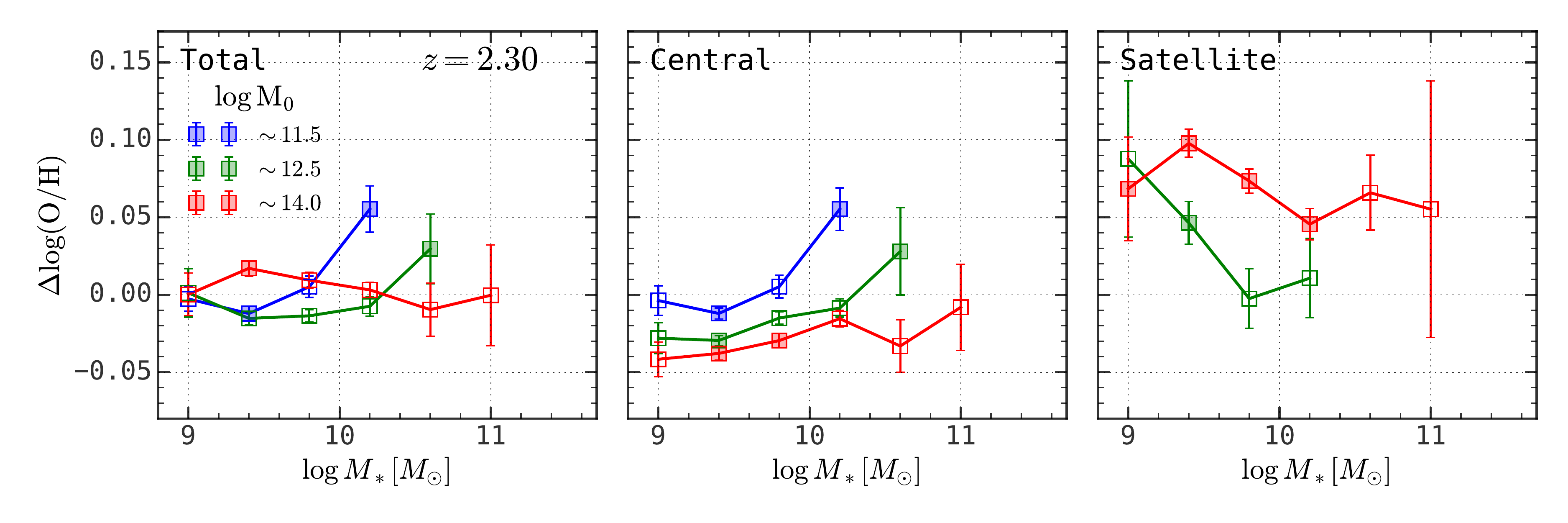}
    \caption{
        Similar to Figure~\ref{fig:figures/mzr_pc_eagle}, except for the TNG100
        simulation.
    }%
    \label{fig:figures/mzr_pc_tng100}
\end{figure*}

Here we use the TNG100 cosmological megneto-hydrodynamical simulation of the
Illustris TNG project \citep{weinbergerSimulatingGalaxyFormation2017,
    pillepichSimulatingGalaxyFormation2018,
    pillepichFirstResultsIllustrisTNG2018,
    marinacciFirstResultsIllustrisTNG2018, nelsonFirstResultsIllustrisTNG2018,
springelFirstResultsIllustrisTNG2018, naimanFirstResultsIllustrisTNG2018}.
It simulates galaxy formation and evolution in a box with a side length of
100cMpc with $1820^3$ dark matter particles and $1820^3$ baryonic
particles. Each dark matter particle weighs $\sim 7.5\times 10^6M_{\odot}$
and each baryonic particle weighs $\sim 1.4\times 10^6M_{\odot}$.

Dark matter halos are identified with the FoF algorithm
Subhalos and galaxies are identified with the \texttt{SUBFIND} algorithm. In
each FoF halo, the subhalo with the lowest gravitational potential is defined
as the central subhalo and the galaxy in it is defined as the central galaxy,
while others are satellite galaxies.

Stellar mass is defined as the total mass of star particles within twice the
half stellar mass radius, $R_*$, where $R_*$ is the radius that enclosed half
of the stellar mass in each subhalo. The gas phase metallicity is defined as
the abundance ratio between oxygen and hydrogen for all gas particles within
$2R_*$.

The top panels in Figure~\ref{fig:figures/mzr_bin_z_tng100} show the gas phase
metallicity difference for galaxies with different host halos, while the middle
and bottom panels show the results for central and satellite galaxies
separately. Central galaxies in massive halos are more metal-poor than those in
low-mass halos. On the contrary, satellite galaxies in massive halos are
more metal-rich than those in low-mass ones. Despite different subgrid recipes
adopted in TNG and EAGLE, the environmental dependence of the gas phase
metallicity agree with each other quite well.

Finally, Figure~\ref{fig:figures/mzr_pc_tng100} shows the gas-phase
metallicity as a function of stellar mass for galaxies with different
descendant halo mass, $M_0$. Again, these results agree with those obtained in
the EAGLE simulation quite well.

\section{MZR for central galaxies without AGN feedback}%
\label{sec:mzr_for_central_galaxies_without_agn_feedback}

\begin{figure*}
    \centering
    \includegraphics[width=0.9\linewidth]{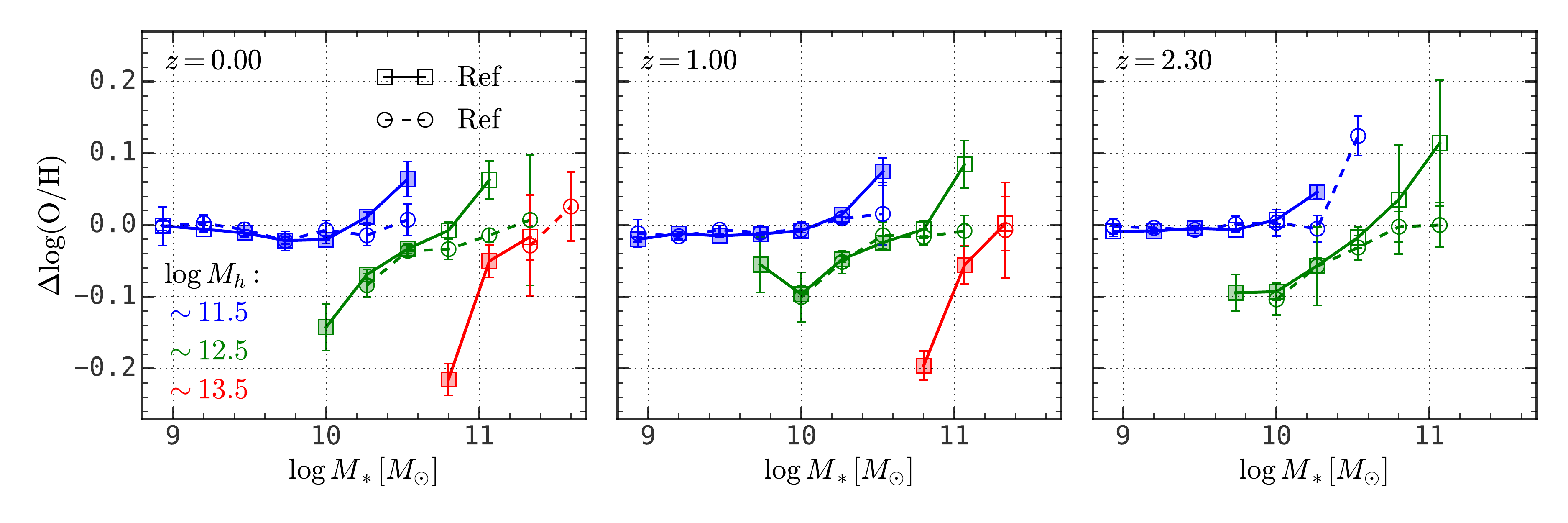}
    \caption{
        Similar to Figure~\ref{fig:figures/mzr_bin_z_eagle}, except for central
        galaxies. The boxes and solid lines are for the reference EAGLE
        simulation, which is labeled as \texttt{Ref-L100N1504}. The circles
        and dashed lines are for the simulation without AGN feedback, which is
        labelled as \texttt{NoAGN-L005N0752}.
    }%
    \label{fig:figures/noagn_eagle_mzr}
\end{figure*}

Figure~\ref{fig:figures/noagn_eagle_mzr} shows the MZR difference of central
galaxies in halos with different mass for the reference EAGLE simulation and
the one with AGN feedback disabled. Here one can see that the gas phase
metallicity difference for central galaxies in different mass of halos is
reduced when the AGN feedback is disabled, especially for massive galaxies at
low $z$. Meanwhile, the MZR difference at high $z$ is not affected.

\end{document}